\documentclass[fleqn,11pt]{article}

\usepackage{amsmath,amsfonts,amssymb}

\usepackage{geometry}
\newtheorem{theorem}{Theorem}
\newtheorem{lemma}[theorem]{Lemma}

\newcommand{\R}{\mathbb{R}}

\newcommand{\N}{\mathbb{N}}

\newcommand{\cqfd}
{%
\mbox{}%
\nolinebreak%
\hfill%
\rule{2mm}{2mm}%
\newline
\newline
}

\title{A linearized kinetic problem on the half-line with collision operator from a Bose condensate with excitations.}
\author{Leif ARKERYD and Anne NOURI\\
\\Mathematical Sciences, 41296 G\"oteborg, Sweden,\\
LATP, Aix-Marseille University, France}

\date{}

\begin{document}

\maketitle

{\noindent \bf Abstract.}\hspace{0.1in}
This paper deals with a half-space linearized problem for the distribution function of the excitations in a Bose gas close to equilibrium . Existence and uniqueness of the solution, as well as its asymptotic properties are proven for a given energy flow. The problem differs from the ones for the classical Boltzmann and related equations, where the hydrodynamic mass flow along the half-line is constant. Here it is no more constant. Instead we use the energy flow which is constant, but no more hydrodynamic.

\footnotetext[2]{Key words; low temperature kinetics, Bose condensate, two component model, Milne problem.}
%
%
%
%
%
\section{Introduction.}
This paper studies a linearized half-line problem related to the kinetic equation for a gas of excitations interacting with a Bose condensate. Below the temperature $T_c$ where Bose-Einstein condensation sets in, the system consists of a condensate and excitations. The condensate density $n_c$ is modelled by a Gross-Pitaevskii equation. The excitations are described by a kinetic equation with a source term taking into account their interactions with the condensate,
\begin{equation}\label{nonlin-excit}
\frac{\partial F}{\partial t}+p\cdot \nabla _xF= C_{12}(F,n_c).
\end{equation}
With $F$ the distribution function of the excitations, and $n_c$ the density of the condensate, the collision operator in this model is
\begin{equation}\label{C12}
C_{12}(F,n_c)(p)= n_c\int \delta _0\delta _3\Big( (1+F_1)F_2F_3-F_1(1+F_2)(1+F_3)\Big) dp_1dp_2dp_3,
\end{equation}
where $F(p_i)$ is denoted by $F_i$, and
\[ \begin{aligned}
&\delta _0= \delta (p_1= p_2+p_3,p_1^2= p_2^2+p_3^2+n_c),\quad \delta _3= \delta ( p_1=p)-\delta ( p_2=p)-\delta ( p_3=p).
\end{aligned}\]
This corresponds to the 'high temperature case' $|p|\gg \sqrt{n_c}$ in the superfluid rest frame with the temperature range close to $0.7\hspace{.02cm} T_c$, where the approximation $p^2+n_c$ for the excitation energy is commonly used.\\
Multiplying (1.2) by $\log \frac{F}{1+F}$ and integrating in p, it follows that $C_{12}(F,n_c)= 0$ if and only if
 \begin{equation*}
\frac{F_1}{1+F_1}=\frac{F_2}{1+F_2}\frac{F_3}{1+F_3}, \quad p_1= p_2+p_3,p_1^2= p_2^2+p_3^2+n_c.
 \end{equation*}
This implies that the kernel of $C_{12}$ consists of the Planckian distribution functions
\begin{eqnarray*}
P_{\alpha ,\beta }(p)= \frac{1}{e^{\alpha (p^2+n_c)+\beta \cdot p}-1},\quad p\in \R^3,\quad \text{for}\quad \alpha >0,\hspace*{0.05in}\beta \in \R^3.
\end{eqnarray*}
We refer to \cite{AN1} and references therein for a further discussion of the two-component model, and to \cite{AN3} where its well-posedness and long time behaviour are studied close to equilibrium. In that context the linearized half-space problem of this paper is connected to boundary layer questions for (\ref{nonlin-excit}), for which $n_c$ may be taken as a constant $n$. Take $\alpha=1$ and write $(|p|^2+n)+\beta\cdot p=|p+\frac{\beta}{2}|^2+n-\frac{|\beta|^2}{4}$. With the approximation (close to diffusive thermal equilibrium) $n-\frac{|\beta|^2}{4}=0$, i.e. $|\beta|=2\sqrt{n}$, the Planckian $P(p)$ takes the form
\begin{eqnarray*}
P(p)=\frac{1}{e^{|p-p_0|^2}-1}\quad {\rm with}\quad p_0=\frac{\beta}{2}.
\end{eqnarray*}
Changing variables $p-p_0\rightarrow p$ gives
\begin{eqnarray*}
P(p)=\frac{1}{e^{|p|^2}-1}.
\end{eqnarray*}
The Dirac measure $\delta_0$ in (1.2) changes into $\delta_c=\delta(p_1=p_2+p_3+p_0, p_1^2=p_2^2+p_3^2)$.
\[\]
With $F=P(1+f)$, the integrand of the collision operator becomes
\[ \begin{aligned}
(1+F_1)F_2F_3-F_1(1+F_2)(1+F_3)= &-(1+P_{2}+P_{3})P_{1}f_1+(P_{3}-P_{1})P_{2}f_2+(P_{2}-P_{1})P_{3}f_3\\
&+P_{2}P_{3}f_2f_3-P_{1}P_{2}f_1f_2-P_{1}P_{3}f_1f_3.
\end{aligned}\]
Here we have used that $(1+P)=M^{-1}P$, where
\begin{eqnarray*}
M(p)= e^{-p^2},\hspace*{0.03in} p\in \R^3,
\end{eqnarray*}
and that $M(p_1)= M(p_2)M(p_3)$ when $p_1^2= p_2^2+p_3^2$. It follows that the linear term in the previous integrand gives the linearized operator
\[ \begin{aligned}
L(f)=\frac{ n}{P}\int \delta _c\delta _3
\Big[  &-(1+P_{2}+P_{3})P_{1}f_1+(P_{3}-P_{1})P_{2}f_2+(P_{2}-P_{1})P_{3}f_3\Big]dp_1dp_2dp_3 .
\end{aligned}\]
We shall here consider functions on a half-line in the $x$-direction, which in the variable $p=(p_x,p_y,p_z)$ are cylindrically symmetric functions of $p_x$ and $p_r=\sqrt{p_y^2+p_z^2}$. Assuming $p_0=(0,p_{0y},p_{0z})$, this changes the momentum conservation Dirac measure in $L$ to $\delta(p_{1x}-p_{2x}-p_{3x})$. Being in the high temperature case, we introduce a cut-off at $\lambda>0$ in the integrand of $L$, given by the characteristic function $\tilde{\chi}$ for the set of $(p,p_1,p_2,p_3)$, such that
 \begin{equation*}
|p|\geq \lambda,\quad |p_1|\geq\lambda,\quad |p_2|\geq\lambda,\quad |p_3|\geq\lambda.
 \end{equation*}
The Milne problem is
\begin{equation}\label{milne1}
p_x\partial _xf= Lf, \quad x>0,\hspace*{0.05in}p_x\in \R,\hspace*{0.05in}p_r\in \R ^+,\hspace{0.05in}|p|\geq\lambda,
\end{equation}
\begin{equation}\label{milne-bc1}
f(0,p)= f_0(p_x,p_r),\quad p_x>0,\quad |p|\geq\lambda,
\end{equation}
where $f_0$ is given. The restriction $|p|\geq\lambda$ will be implicitly assumed below, and $\int dp$ will stand for $\int_{|p|\geq\lambda}dp$. \\
We shall prove in Section 2 (see (2.1)) that the kernel of $ L$ is spanned by $|p|^2(1+P)$ and $p_x(1+P)$.
For any measurable function $f(x,p)$ such that for almost all $x\in \R^+$,
\begin{eqnarray*}
\Big( p\rightarrow f(x,p)\Big) \in L^2_{p_r(1+|p|)^3\frac{P}{1+P}}(\R \times \R^+),
\end{eqnarray*}
where $|p|=\sqrt{p_x^2+p_r^2}$, let
\begin{equation}\label{ortho-decomp}
f(x,p)= a(x)|p|^2(1+P)+b(x)p_x(1+P)+w(x,p)
\end{equation}
be its orthogonal decomposition on the kernel of $L$ and the orthogonal complement in $L^2_{p_r\frac{P}{1+P}}$, i.e.
\begin{equation}\label{ortho-cond}
\int p_xw(x,p)Pp_rdp_xdp_r= \int |p|^2w(x,p)Pp_rdp_xdp_r= 0,\quad x\in \R^+.
\end{equation}
Denote by $D$ the function space
\[ \begin{aligned}
&D= \{ f; \hspace*{0.03in}f\in L^\infty (\R^+; L^2_{p_r(1+|p|)^3\frac{P}{1+P}}(\R \times \R^+)),\hspace*{0.03in} p_x\partial _x f\in L^2_{loc}(\R^+; L^2_{p_r(1+|p|)^{-3}\frac{P}{1+P}}(\R \times \R^+))\} .\\
\end{aligned}\]
The main result of this paper is the following.
%
%
%
%
%
\begin{theorem}\label{th-milne}
For any $\mathcal{E}\in \R$ and $f_0\in L^2_{p_r(1+\lvert p\rvert )^3\frac{P}{1+P}}(\R^+\times \R ^+)$, there is a unique solution $f\in D$ to the Milne problem,
\begin{equation}\label{milne1}
p_x\partial _xf= Lf, \quad x>0,\hspace*{0.05in}p_x\in \R,\hspace*{0.05in}p_r\in \R ^+,
\end{equation}
\begin{equation}\label{milne-bc1}
f(0,p)= f_0(p),\quad p_x>0,
\end{equation}
\begin{equation}\label{energy-flow1}
\int p_x |p|^2 f(x,p)P(p)dp= \mathcal{E},\quad x\in \R ^+.
\end{equation}
Moreover, for the decomposition (\ref{ortho-decomp}) of the solution, there are $(a_\infty ,b_\infty )\in \R ^2$ with
\begin{equation}\label{cond-infinity-ab}
b_\infty =\frac{\mathcal{E}}{\gamma},\quad{\rm where}\quad \gamma = \int p_x^2 |p|^2 P(1+P)dp,
\end{equation}
and a constant $c>0$, such that for any $\eta \in ] 0, c_1[$,
\begin{equation}\label{exp-cv}
\int (1+|p|)^3w^2(x,p)\frac{P}{1+P}dp+|a(x)-a_\infty |^2+|b(x)-b_\infty |^2\leq ce^{-2\eta x},\quad x\in \R ^+.
\end{equation}
Here with $\nu_0$ defined by (2.4),
\begin{equation}\label{df-c4}
c_1= \min \{ \frac{\nu _0}{2},\frac{\nu _0}{2c_2}\},\quad c_2= \frac{2}{\gamma }\Big( \int p_x^4P(1+P)dp\int p_x^2 |p|^4 P(1+P)dp\Big) ^{\frac{1}{2}}.
\end{equation}
\end{theorem}
\hspace*{0.2in}\\
{\bf Remarks.}\\
{This result should be compared to the analogous result concerning the Milne problem for the linearized Boltzmann operator around the absolute Maxwellian in \cite{BCN}. In  \cite{BCN}  the mass flow is constant and well-posedness for the Milne problem is proven for a given mass flow. In the present paper on the other hand, the mass flow may not be constant, since mass is not a hydrodynamic mode. But the energy flow is constant, and well-posedness here is proven for a fixed energy flow.
That this energy flow is proportional to the asymptotic limit of the mass flow, is a new low temperature result.\\
A separate complication in the present case is that, whereas the given mass flow in \cite{BCN} is a hydrodynamic component of the solution, here the energy flow is not in the kernel of $L$.
Another differing aspect compared to classical kinetic theory, is that the collision frequency is asymptotically equivalent to $|p|^3$, when $p\rightarrow\infty$.}\\
\hspace*{0.2in}\\
The interest in half space problems such as (\ref{milne1})-(\ref{milne-bc1}) is partly due to their role in the boundary layer behaviour of the solution of boundary-value problems of kinetic equations for small Knudsen numbers. This subject has received much attention for the Boltzmann equation (\cite{M2}, \cite{So1}, \cite{So2}, \cite{UYY1}, \cite{UYY2}, \cite {AN2}) and related equations (\cite{BT}, \cite{P}). Starting from the stationary Boltzmann equation in a half-space with given in-datum and a Maxwellian limit at infinity, the unknown is assumed to stay close to this Maxwellian, giving rise to the linearized stationary Boltzmann equation in a half space. A general treatment of the linearized problem for hard forces and hard spheres under null bulk velocity, is given in \cite{M1} and  references therein. The case of a gas of hard spheres (resp. of hard or soft forces) and a null bulk velocity at infinity is independently treated in \cite{BCN} (resp. in \cite{GP}). The case of a gas of hard spheres and a nonzero bulk velocity at infinity is considered in \cite{CGS}, positively answering a former conjecture \cite{C}. The Milne problem for the Boltzmann equation with a force term is analyzed in \cite{CME}. Half-space problems in a discrete velocity frame are studied in \cite{BB}. For a review of mathematical results on the half-space problem for the linear and nonlinear Boltzmann equations, we refer to \cite{BGS}.\\
\\
The plan of the paper is the following. In Section 2, the linearized collision operator $L$ is studied, including a spectral inequality. In Section 3, Theorem \ref{th-milne} is proven.

%
%
%
%
%
%
%
\setcounter{equation}{0}
\setcounter{theorem}{0}
\section{The linearized collision operator.}
\begin{lemma}
\hspace*{0.2in}\\
$L$ is a self-adjoint operator in $L^2_{\frac{P}{1+P}}$.
Within the space of cylindrically invariant distribution functions,
its kernel is the subspace spanned by $|p|^2(1+P)$ and $p_x(1+P)$.
\end{lemma}
\underline{Proof.} It follows from the equalities
\begin{eqnarray*}
P_2(1+P_2)(P_3-P_1)= P_3(1+P_3)(P_2-P_1)= P_1(1+P_2)(1+P_3),\\
P_1(1+P_2+P_3)= P_2P_3= \frac{P_1(1+P_2)(1+P_3)}{1+P_1},\quad p_1^2=p_2^2+p_3^2,
\end{eqnarray*}
that for any functions $f$ and $g$ in $L^2_{\frac{P}{1+P}}$,
\[ \begin{aligned}
\int \frac{P}{1+P}(p)f(p)Lg(p)dp= &-n\int  \tilde{\chi}\delta _cP_1(1+P_2)(1+P_3)\\
&(\frac{f_1}{1+P_1}-\frac{f_2}{1+P_2}-\frac{f_3}{1+P_3})(\frac{g_1}{1+P_1}-\frac{g_2}{1+P_2}-\frac{g_3}{1+P_3})dp_1dp_2dp_3.
\end{aligned}\]
This proves the self-adjointness of $L$ in $L^2_{\frac{P}{1+P}}$. Moreover, $Lf= 0$ for $f\in L^2_{\frac{P}{1+P}}$ implies that
\begin{eqnarray*}
\frac{f_1}{1+P_1}= \frac{f_2}{1+P_2}+\frac{f_3}{1+P_3},\quad p_{1x}= p_{2x}+p_{3x},\quad p_1^2=p_2^2+p_3^2.
\end{eqnarray*}
It is a consequence of this Cauchy equation that the orthogonal functions
\begin{eqnarray}
|p|^2(1+P) \hspace*{0.05in}\text{and}\hspace*{0.05in}p_x(1+P)
\end{eqnarray}
span the kernel of $L$.\\ \cqfd
\hspace*{0.2in}\\
\hspace{1cm}\\
The operator $L$ splits into $K-\nu$, where
\begin{eqnarray}\label{df-K}
Kf(p):= \frac{2n}{P(p)}\Big( \int \tilde{\chi}\delta (p_x= p_{2x}+p_{3x}, p^2= p_2^2+p_3^2)(P_3-P)P_2f_2dp_2dp_3\nonumber \\
+ \int \tilde{\chi}\delta (p_{1x}= p_x+p_{3x}, p_1^2= p^2+p_3^2)(1+P+P_3)P_1f_1dp_1dp_3\nonumber \\
+\int \tilde{\chi}\delta (p_{1x}=p_x+p_{3x}, p_1^2= p^2+p_3^2)(P_1-P)P_3f_3dp_1dp_3\Big)
\end{eqnarray}
and
\begin{eqnarray}\label{df-nu}
\nu (p):= n\int \tilde{\chi}\delta (p_x= p_{2x}+p_{3x}, p^2= p_2^2+p_3^2)(1+P_2+P_3)dp_2dp_3\nonumber \\
+ 2n\int \tilde{\chi}\delta (p_{1x}= p_x+p_{3x}, p_1^2= p^2+p_3^2)(P_3-P_1)dp_1dp_3.
\end{eqnarray}

\begin{lemma}
The operator $K$ is compact from $L^2_{\nu \frac{P}{1+P}}$ in $L^2_{\nu ^{-1}\frac{P}{1+P}}$. The collision frequency $\nu $ satisfies
\begin{eqnarray}\label{bounds-nu}
\nu _0(1+|p|)^3\leq \nu (p)\leq \nu _1(1+|p|)^3, \quad p= (p_x,p_r)\in \R \times \R^+,
\end{eqnarray}
for some positive constants $\nu _0$ and $\nu_1$.
\end{lemma}
\underline{Proof of Lemma 2.2.}
$K= K_1+K_2+K_3$, where
\[ \begin{aligned}
&K_1h(p):=2\pi n \int k_1(p,p_2)h_2dp_2,\quad K_2h(p)= 2\pi n\int k_2(p,p_1)h_1dp_1,\quad K_3(p)= 2\pi n\int k_3(p,p_3)h_3dp_3,\\
&k_1(p,p_2):= \frac{P_2}{\pi}\int \delta (p_x=p_{2x}+p_{3x},p^2=p_2^2+p_3^2)\frac{P_3-P}{P}dp_3\\
&\hspace*{0.6in}= P_2\hspace*{0.05in}\chi _{p^2-p_2^2-(p_x-p_{2x})^2>0}(\frac{1}{P(e^{p^2-p_2^2}-1)}-1),\\
&k_2(p,p_1):= \frac{P_1}{\pi}\int  \delta (p_{1x}= p_x+p_{3x}, p_1^2= p^2+p_3^2)\frac{1+P+P_3}{P}dp_3\\
&\hspace*{0.6in}= P_1\hspace*{0.05in}\chi _{p_1^2-p^2-(p_{1x}-p_{x})^2>0}(\frac{1}{P}+1+\frac{1}{P(e^{p_1^2-p^2}-1)}),\\
&k_3(p,p_3):= \frac{P_3}{\pi}\int \delta (p_{1x}=p_x+p_{3x}, p_1^2= p^2+p_3^2)\frac{P_1-P}{P}dp_1\\
&\hspace*{0.6in}= P_3\hspace*{0.05in}\chi _{p^2+p_3^2-(p_{x}-p_{3x})^2>0}(\frac{1}{P(e^{p^2+p_3^2}-1)}-1).
\end{aligned}\]
Let $m\in \N^*$.
We treat separately the parts of $K_1$ with $\frac{P_3}{P}$ and with $\frac{P}{P}$, and notice that for $|p|,|p_2|,|p_3|\geq \lambda$, factors $P=\frac{M}{1-M}$ may be repaced by $M$ for questions of boundedness and convergence to zero. For $m>\lambda$ fixed, split the domain of $p_2$ into $|p_2|<m$ and $|p_2|>m$. The mapping
\begin{eqnarray*}
\int_{|p_2|<m} k_{11}(p,p_2)h_2dp_2,\quad \text{where}\quad  k_{11}(p,p_2)=M_2M_3M^{-1}\chi_{p^2-p_2^2-(p_x-p_{2x})^2>  0,\lambda^2<|p_2|^2<p^2-\lambda^2},
\end{eqnarray*}
is compact from $L^2_{\nu\frac{P}{1+P}}$ into $L^2_{\nu^{-1}\frac{P}{1+P}}$. Indeed
\[\begin {aligned}
&\int_{|p_2|<m}\nu^{-1} Mk_{11}^2(p,p_2)\nu_2 ^{-1}M_2^{-1}dpdp_2<\infty.
\end{aligned}\]
The mapping $h\rightarrow \int_{|p_2|>m}k_{11}(p,p_2)h_2dp_2$ tends to zero in $L^2_{\nu^{-1}\frac{P}{1+P}}$ when $m\rightarrow\infty$, uniformly for functions $h$ with norm one  in $L^2_{\nu\frac{P}{1+P}}$. Namely, it holds
\[\begin{aligned}
(\int \nu^{-1}M(\int_{|p_2|>m}k_{11}(p,p_2)h_2dp_2)^2dp)^{\frac{1}{2}}&\leq
\int_{|p_2|>m}(\int \nu^{-1}Mk_{11}^2(p,p_2)dp)^{\frac{1}{2}}h_2dp_2\\
\leq \int _{\lvert p_2\rvert >m}(\int _{\lvert p\rvert >\lvert p_2\rvert }\nu ^{-1}Mdp)^{\frac{1}{2}}h_2dp_2
&\leq c\int _{\lvert p_2\rvert >m}\frac{M_2^{\frac{1}{2}}}{\lvert p_2\rvert }h_2dp_2
\leq \frac{c}{m}(\int M_2\nu _2h_2^2dp_2)^{\frac{1}{2}}.
\end{aligned}\]
The other term in $K_1$ only differs in the factor $\frac{P}{P} < \frac{P_3}{P}$. The compactness of $K_1$ follows.\\
An analogous splitting of $K_2$ with respect to velocities
smaller and larger than $m$, gives for $K_2$ and
$|p_1|<m$ that the dominating term corresponds to the factor $\frac{1}{P}$.
The mapping becomes
\begin{eqnarray*}
h\rightarrow \int_{|p_1|<m} k_{21}(p,p_1)h_1dp_1,\quad \text{with}\quad k_{21}(p,p_1)=M_1M^{-1}\chi_{p_1^2-p^2-(p_{1x}-p_{x})^2>  0,\lambda^2<p_1^2-p^2}.
\end{eqnarray*}
Since the kernel $k_{21}$ is bounded
on the domain of integration which is bounded, this mapping is compact.
The mapping $h\rightarrow \int_{|p_1|>m}k_{21}(p,p_1)h_1dp_1$ tends to zero in $L^2_{\nu^{-1}\frac{P}{1+P}}$ when $m\rightarrow\infty$, uniformly for functions $h$ with norm one  in $L^2_{\nu\frac{P}{1+P}}$. Here
\[ \begin{aligned}
(\int \nu^{-1}M(\int_{|p_1|>m}k_{21}(p,p_1)h_1dp_1)^2dp)^{\frac{1}{2}}&\leq
\int_{|p_1|>m}(\int_{p^2<p_1^2} \nu^{-1}M^{-1}dp)^{\frac{1}{2}}M_1^{\frac{1}{2}}\nu _1^{-\frac{1}{2}}(M_1\nu_1)^{\frac{1}{2}}h_1dp_1\\
&\leq c\int_{|p_1|>m}\nu_1^{-\frac{5}{6}}(M_1\nu_1)^{\frac{1}{2}}h_1dp_1\\
&\leq
(\int_{|p_1|>m}\nu_1^{-\frac{5}{3}}dp_1)^{\frac{1}{2}}(\int M_1\nu_1h_1^2dp_2)^{\frac{1}{2}},
\end{aligned}\]
which again tends to zero, uniformly in $h$ when $ m\rightarrow \infty$.
In $K_3$ the dominating term corresponds to the factor  $\frac{P}{P}$. For the kernel
\[\begin {aligned}
k_{31}(p,p_3)= M_3 \chi_{p^2+p_3^2-(p_x+p_{3x})^2>0,|p_3|>\lambda},
\end{aligned}\]
it holds that
\[\begin {aligned}
\int_{|p_3|<m}\nu^{-1} Mk_{31}^2(p,p_3)\nu_3 ^{-1}M_3^{-1}dpdp_3<\infty,
\end{aligned}\]
and
\[ \begin{aligned}
(\int \nu^{-1}M(\int_{|p_3|>m}k_{31}(p,p_3)h_3dp_3)^2dp)^{\frac{1}{2}}&\leq
(\int_{|p_3|>m}(\int \nu^{-1}Mk_{31}^2(p,p_3)dp)^{\frac{1}{2}}h_3dp_3\\
&\leq \int_{|p_3|>m} M_3h_3dp_3(\int \nu ^{-1}Mdp)^{\frac{1}{2}}\\
&\leq
c(\int_{|p_3|>m}M_3\nu _3^{-1}dp_3)^{\frac{1}{2}}(\int M_3\nu_3h_3^2dp_3)^{\frac{1}{2}}.
\end{aligned}\]
This ends the proof of the compactness of $K$.\\
\hspace{1cm}\\
The function $\nu $ is bounded from below by a positive constant, since
\begin{eqnarray*}
P_3-P_1>0,\quad p^2_1= p^2+p^2_3.
\end{eqnarray*}
\hspace*{0.2in}\\
For $|p|>\lambda $ the first term of $\nu (p)$ belongs to the interval with end points
\[ \begin{aligned}
&2\pi ^2n\int _{p_{2r}>0,\hspace*{0.01in} p_{2r}^2+2(p_{2x}-\frac{1}{2}p_x)^2<\frac{1}{2}p_x^2+p_r^2 }p_{2r}dp_{2r}dp_{2x}\\
&\text{and}\\
&2\pi ^2n(1+\frac{2}{e^{\lambda ^2}-1})\int _{p_{2r}>0, \hspace*{0.01in} p_{2r}^2+2(p_{2x}-\frac{1}{2}p_x)^2<\frac{1}{2}p_x^2+p_r^2 }p_{2r}dp_{2r}dp_{2x}.
\end{aligned}\]
With the change of variables $(x,y):= (p_{2x}, p_{2r}^2)$,
\[ \begin{aligned}
&2\int _{p_{2r}>0, \hspace*{0.01in} p_{2r}^2+2(p_{2x}-\frac{1}{2}p_x)^2
<\frac{1}{2}p_x^2+p_r^2 }p_{2r}dp_{2r}dp_{2x}\\
&= \int _{y>0,\hspace*{0.01in}  y+2(x-\frac{1}{2}p_x)^2<\frac{1}{2}p_x^2+p_r^2}dxdy\\
&= \int _0^{\frac{1}{2}p_x^2+p_r^2}\int _{(x-\frac{1}{2}p_x)^2<\frac{1}{4}(p_x^2+2p_r^2-2y)}dxdy\\
&= \int _0^{\frac{1}{2}p_x^2+p_r^2}\sqrt{p_x^2+2p_r^2-2y}dy\\
&= \frac{1}{3}(p_x^2+2p_r^2)^{\frac{3}{2}}\\
&\sim |p|^3.
\end{aligned}\]
The second term of $\nu (p)$ is bounded. Indeed,
\[ \begin{aligned}
0&\leq\frac{1}{\pi^2} \int \delta (p_{1x}= p_x+p_{3x}, p_1^2= p^2+p_3^2)(P_3-P_1)dp_1dp_3\\
&\leq 2\int _{p_{1r}>0}(\frac{1}{e^{p_1^2-p^2}-1}-P_1)\Big( \int _0^{+\infty }\delta (p_{3r}^2= p_{1r}^2-p^2-p_x^2+2p_xp_{1x})p_{3r}dp_{3r}\Big) p_{1r}dp_{1r}dp_{1x}\\
&\leq \iint _0^{+\infty }\frac{1}{e^{x^2+y-p^2}-1}\chi _{y>p^2+p_x^2-2xp_x}dydx\\
&= \sum _{k\geq 1}e^{kp^2}\int e^{-kx^2}\int _0^{+\infty }e^{-ky}\chi _{y>p^2+p_x^2-2xp_x}dydx\\
&\leq \sum _{k\geq 1}\frac{1}{k}\int e^{-k(x-p_x)^2}dx\\
&= \sqrt{\pi }\sum _{k\geq 1}\frac{1}{k^{\frac{3}{2}}}.
\end{aligned}\]
\cqfd

\hspace{1cm}\\
Denote by $(\cdot , \cdot )$ the scalar product in $L^2_{\frac{P}{1+P}}$, and by $\tilde{P}$ the orthogonal projection on the kernel of $L$.
\begin{lemma}
\hspace*{0.2in}\\
$L$ satisfies the spectral inequality,
\begin{equation}\label{spectral-in}
- (Lf, f)\geq \nu _0((1+|p|)^3(I-\tilde{P})f,(I-\tilde{P})f),\quad f\in L^2_{(1+\lvert p\rvert )^3\frac{P}{1+P}}.
\end{equation}
\end{lemma}
{\bf Proof.}\hspace*{0.03in}For the compact, self-adjoint operator $K$, the spectrum behaves similarly to the classical Boltzmann case. Namely, there is no eigenvalue $\alpha>1$ for $\frac{K}{\nu}$. Else there is $f\neq 0$ such that $Lf= (\alpha-1)\nu f$ and so $(Lf,f)>0$. But
\begin{eqnarray*}
(Lf,f)= -n\int \tilde{\chi}\delta _c(\frac{f_1}{1+P_1}-\frac{f_2}{1+P_2}-\frac{f_3}{1+P_3})^2dp_1dp_2dp_3\leq 0.
\end{eqnarray*}
 In the complement of the kernel of $L$, the eigenvalues of $\frac{K}{\nu}$ are bounded from above by $\alpha_0<1$. Spanning $L^2_{(1+\lvert p\rvert )^3\frac{P}{1+P}}$ with the coresponding eigenfunctions of $\frac{K}{\nu}$ and  the kernel of $L$, we obtain the spectral inequality
\begin{eqnarray*}
(Lf,f)\leq (\alpha_0-1)(\nu (I-\tilde{P})f,(I-\tilde{P})f),\quad f\in L^2_{(1+\lvert p\rvert )^3\frac{P}{1+P}}.
\end{eqnarray*}
From here, (\ref{spectral-in}) follows by (\ref{bounds-nu}).\\
%
%
%
%
%
%
%
%
\setcounter{equation}{0}
\section{The Milne problem}
This section gives the proof of Theorem \ref{th-milne}. \\
Let
\begin{eqnarray*}
\tilde{f}= f-\frac{\mathcal{E}}{\gamma }p_x(1+P),\quad \tilde{f}_0(p)= f_0(p)-\frac{\mathcal{E}}{\gamma }p_x(1+P).
\end{eqnarray*}
Solving the Milne problem (\ref{milne1})-(\ref{milne-bc1})-(\ref{energy-flow1}) for the unknown $f$ is equivalent to solving
\begin{equation}\label{milne2}
p_x\partial _x\tilde{f}= L\tilde{f}, \quad x>0,\hspace*{0.05in}p_x\in \R,\hspace*{0.05in}p_r\in \R ^+,
\end{equation}
\begin{equation}\label{milne-bc2}
\tilde{f}(0,p)= \tilde{f}_0(p),\quad p_x>0,
\end{equation}
\begin{equation}\label{energy-flow2}
\int p_x|p|^2\tilde{f}(x,p)P(p)dp= 0,\quad x\in \R ^+,
\end{equation}
for the unknown $\tilde{f}$.\\
\hspace*{0.2in}\\
%
%
%
%
%
%
We first study the behaviour of a solution $\tilde{f}$ to the Milne problem (\ref{milne2})-(\ref{milne-bc2})-(\ref{energy-flow2}), when $x\rightarrow +\infty $.\\
Set
\begin{eqnarray*}
\tilde{f}(x,p)= (a(x)|p|^2+\tilde{b}(x)p_x)(1+P)+w(x,p),
\end{eqnarray*}
with
\begin{eqnarray*}
\int p_xwPdp= \int |p|^2wPdp= 0,
\end{eqnarray*}
an orthogonal decomposition of $\tilde{f}$. Denote by
\begin{equation}\label{df-W}
W(x)= \frac{1}{2}\int p_x\tilde{f}^2(x,p)\frac{P}{1+P}dp
\end{equation}
the linearized entropy flux of $\tilde{f}$. It holds that
\begin{equation}\label{W0}
W(0)\leq \frac{1}{2}\int _{p_x> 0}p_x\tilde{f}_0^2(p)\frac{P}{1+P}dp.
\end{equation}
By (3.3)
\[ \begin{aligned}
W(x)&= \frac{1}{2}\int p_x\tilde{f}^2(x,p)\frac{P}{1+P}dp-\frac{1}{\gamma }\int p_x^2\tilde{f}(x,p)P(p)dp\int p_x|p|^2\tilde{f}(x,p)P(p)dp\\
&=  \frac{1}{2}\int p_xw^2(x,p)\frac{P}{1+P}dp+a\int p_x|p|^2wPdp+\tilde{b}\int p_x^2wPdp+a\tilde{b}\gamma \\
&-\frac{1}{\gamma }\Big( \int p_x^2wPdp+a\gamma \Big) \Big( \int p_x|p|^2wPdp+\tilde{b}\gamma \Big) ,
\end{aligned}\]
i.e.
\begin{equation}\label{W-w}
W(x)=  \frac{1}{2}\int p_xw^2(x,p)\frac{P}{1+P}dp-\frac{1}{\gamma }\int p_x^2wPdp\int p_x|p|^2wPdp.
\end{equation}
This differs from ([BCN]), where the linearized entropy flux of the solution is equal to the linearized entropy flux of its non-hydrodynamic component.The expression (3.6) for $W$ in terms of $w$ is important in the proof.\\
\hspace*{0.2in}\\
Multiplying (\ref{milne2}) by $\tilde{f}\frac{P}{1+P}$, integrating on $(0,X)\times \R \times \R ^+$ (resp. $\R \times \R ^+$), and using (\ref{spectral-in}), gives
\begin{equation}\label{integrated-w}
W(X)+\nu _0\int _0^X\int (1+|p|)^3w^2(x,p)\frac{P}{1+P}dpdx\leq W(0),\quad X>0,
\end{equation}
and
\begin{equation}\label{pointwise-w}
W^\prime (x)+\nu _0\int (1+|p|)^3w^2(x,p)\frac{P}{1+P}dp\leq 0.
\end{equation}
Since $\tilde{f}\in D$, it holds that $W\in L^\infty (\R ^+)$. Then by (\ref{integrated-w}) and (\ref{W-w}), $W\in L^1(\R ^+)$. By (\ref{pointwise-w}), $W$ is a non-increasing function. Hence it tends to zero, when $x$ tends to $+\infty $ and is a nonnegative function. Let $\eta \in ] 0, c_1[$. Multiply (\ref{pointwise-w}) by $e^{2\eta x}$, so that
\begin{eqnarray*}
(W(x)e^{2\eta x})^\prime -2\eta W(x)e^{2\eta x}+\nu _0e^{2\eta x}\int (1+|p|)^3w^2(x,p)\frac{P}{1+P}dp\leq 0.
\end{eqnarray*}
By the Cauchy-Schwartz inequality,
\begin{eqnarray*}
| \int p_x^2w(x,p)Pdp\int p_x|p|^2w(x,p)Pdp |\leq \frac{\gamma c_3}{2}\int w^2(x,p)\frac{P}{1+P}dp.
\end{eqnarray*}
Hence,
\begin{equation}\label{W-1}
(W(x)e^{2\eta x})^\prime +e^{2\eta x}\int \Big( \nu _0(1+|p|)^3-\eta (p_x+c_3)\Big) w^2(x,p)\frac{P}{1+P}dp\leq 0,\quad x\geq 0.
\end{equation}
By the definition (\ref{df-c4}) of $c_1$, the nonnegativity of $W$ and (\ref{W0}), it holds that
\begin{equation}\label{exp-decay-w}
\int _0^\infty e^{2\eta x}\int (1+|p|)^3w^2(x,p)\frac{P}{1+P}dpdx\leq c,
\end{equation}
for some constant $c$. Moreover, by (\ref{W-1}) and (\ref{pointwise-w}),
\begin{equation}\label{exp-decrease-W}
0\leq W(x)\leq W(0)e^{-2\eta x}\leq ce^{-2\eta x},\quad x\geq 0.
\end{equation}
(\ref{exp-decay-w}) implies that $\tilde{f}(x,\cdot )$ converges to a hydrodynamic state when $x\rightarrow +\infty $. In order to prove the exponential point-wise decay of $\int (1+|p|)^3w^2(x,p)\frac{P}{1+P}dp$ in (\ref{exp-cv}), let $0<Y<X$ be given and introduce a smooth cutoff function $\Phi (x)$ such that
\begin{eqnarray*}
\Phi (x)= 0, \hspace*{0.04in}x\in [ 0, \frac{Y}{2}[ \hspace*{0.04in}\cup \hspace*{0.04in}] X+1,+\infty [ ,\quad \Phi (x)= 1,\hspace*{0.04in}x\in [ Y,X] .
\end{eqnarray*}
Denote by $\varphi (x)= e^{\eta x}\Phi (x)$. Then,
\begin{equation}\label{milne-derivative}
p_x\partial _x^2(\varphi \tilde{f})= L(\partial_x(\varphi \tilde{f}))+\varphi ^\prime Lw+p_x\varphi ^{\prime \prime }\tilde{f}.
\end{equation}
Multiply (\ref{milne-derivative}) by $\partial _x(\varphi \tilde{f}) \frac{P}{1+P}$, integrate over $\R _{p_x}\times \R ^+_{p_r}$  and use (\ref{spectral-in}). Hence,
\[ \begin{aligned}
&\frac{1}{2}\frac{d}{dx}\int p_x(\partial _x(\varphi \tilde{f}))^2\frac{P}{1+P}dp+\nu _0\int (1+|p|)^3(\partial _x(\varphi w))^2\frac{P}{1+P}dp\\
&\leq \int \partial _x(\varphi \tilde{f})\Big( \varphi ^\prime Lw+p_x\varphi ^{\prime \prime }\tilde{f}\Big) \frac{P}{1+P}dp\\
&= \varphi ^\prime \int \partial _x(\varphi w)Lw\frac{P}{1+P}dp+\varphi \varphi ^{\prime \prime }\int p_x\tilde{f}\partial _x\tilde{f}\frac{P}{1+P}dp+\varphi ^\prime \varphi ^{\prime \prime }\int p_x\tilde{f}^2\frac{P}{1+P}dp,
\end{aligned}\]
i.e.
\begin{eqnarray*}
\frac{1}{2}\frac{d}{dx}\int p_x(\partial _x(\varphi \tilde{f}))^2\frac{P}{1+P}dp+\nu _0\int (1+|p|)^3(\partial _x(\varphi w))^2\frac{P}{1+P}dp\\
\leq \varphi ^\prime \int \partial _x(\varphi w)Lw\frac{P}{1+P}dp+(\varphi \varphi ^{\prime \prime }W)^\prime +(\varphi ^\prime \varphi ^{\prime \prime }-\varphi \varphi ^{(3)})W.
\end{eqnarray*}
Integrate the last inequality on $[ 0, +\infty[ $, so that
\[ \begin{aligned}
&\nu _0\int _0^{+\infty }\int (1+|p|)^3(\partial _x(\varphi w))^2\frac{P}{1+P}dpdx\\
&\leq \int _0^{+\infty }\varphi ^\prime (x)\int \partial _x(\varphi w)Lw\frac{P}{1+P}dpdx+\int _0^{+\infty }(\varphi ^\prime \varphi ^{\prime \prime }-\varphi \varphi ^{(3)})W(x)dx\\
&\leq \parallel \varphi ^\prime \parallel _{\infty }^2\frac{\alpha }{2}\int _0^{+\infty }\int (1+\lvert p\rvert )^3(\partial _x(\varphi w))^2\frac{P}{1+P}dpdx+\frac{1}{2\alpha }\int _0^{+\infty }\int \frac{1}{(1+\lvert p\rvert )^3}(Lw)^2\frac{P}{1+P}dpdx\\
&+\int _0^{+\infty }(\varphi ^\prime \varphi ^{\prime \prime }-\varphi \varphi ^{(3)})W(x)dx\\
&\leq \parallel \varphi ^\prime \parallel _{\infty }^2\frac{\alpha }{2}\int _0^{+\infty }\int (1+\lvert p\rvert )^3(\partial _x(\varphi w))^2\frac{P}{1+P}dpdx+\frac{c}{2\alpha }\int _0^{+\infty }\int (1+\lvert p\rvert )^3w^2\frac{P}{1+P}dpdx\\
&+\int _0^{+\infty }(\varphi ^\prime \varphi ^{\prime \prime }-\varphi \varphi ^{(3)})W(x)dx,\quad \alpha >0.
\end{aligned}\]
Choose $\alpha <\frac{\nu _0}{\parallel \varphi ^\prime \parallel _{\infty }^2}$. Use (\ref{exp-decay-w}), the Cauchy-Schwartz inequality, and the exponential decay of $W$ expressed in (\ref{exp-decrease-W}) in the W-term. It then holds
\[ \begin{aligned}
\int _0^{+\infty }\int (1+|p|)^3(\partial _x(\varphi w))^2\frac{P}{1+P}dpdx\leq c,
\end{aligned}\]
for some positive constant $c$. Finally,
\begin{equation}\label{pointwise-exp-w-decay}
e^{2\eta X}\int  (1+|p|)^3w^2(X,p)\frac{P}{1+P}dp= 2\int _0^X\int (1+|p|)^3(\partial _x(\varphi w))^2\frac{P}{1+P}dpdx\leq c,\quad X>0.
\end{equation}
\hspace*{0.2in}\\
%
%
%
%
%
%
The exponential decay of $(a,\tilde{b})$ to some limit $(a_\infty ,\tilde{b}_\infty )$ when $x$ tends to $+\infty $, can be proved as follows. The solution $\tilde{f}(x,p)= (a(x)|p|^2+\tilde{b}(x)p_x)(1+P)+w(x,p)$ is solution to (\ref{milne2}) if and only if
\begin{eqnarray*}
(a^\prime p_x|p|^2+\tilde{b}^\prime p_x^2)(1+P)+p_x\partial _xw= Lw.
\end{eqnarray*}
Multiply the former equation by $p_xP$ (resp. $|p|^2P$) and integrate with respect to $p$, so that
\begin{eqnarray*}
(a+\frac{1}{\gamma }\int p_x^2w(\cdot ,p)Pdp)^\prime = (\tilde{b}+\frac{1}{\gamma }\int p_x|p|^2w(\cdot ,p)Pdp)^\prime = 0.
\end{eqnarray*}
Denote by
\begin{eqnarray*}
a_\infty := a(0)+\frac{1}{\gamma }\int p_x^2w(0 ,p)Pdp,\quad \tilde{b}_\infty := \tilde{b}(0)+\frac{1}{\gamma }\int p_x|p|^2w(0 ,p)Pdp.
\end{eqnarray*}
By the Cauchy-Schwartz inequality and (\ref{pointwise-exp-w-decay}),
\begin{eqnarray*}
&|a(x)-a_\infty |= \frac{1}{\gamma }|\int p_x^2w(x ,p)Pdp|\leq c\Big( \int (1+|p|)^3w^2(x,p)\frac{P}{1+P}dp\Big) ^{\frac{1}{2}}\leq ce^{-\eta x},
\end{eqnarray*}
\begin{equation}\label{asympt-b}
|\tilde{b}(x)-\tilde{b}_\infty |= \frac{1}{\gamma }|\int p_x|p|^2w(x ,p)Pdp|\leq c\Big( \int (1+|p|)^3w^2(x,p)\frac{P}{1+P}dp\Big) ^{\frac{1}{2}}\leq ce^{-\eta x}.
\end{equation}
\hspace*{0.2in}\\
%
%
%
By (\ref{exp-decrease-W})
\begin{equation}\label{asympt-uniq1}
\lim _{x\rightarrow +\infty }\int p_x\tilde{f}^2(x,p)\frac{P}{1+P}dp= 0.
\end{equation}
By (\ref{pointwise-exp-w-decay}))
\begin{equation}\label{asympt-uniq2}
\lim _{x\rightarrow +\infty }\int  (1+|p|)^3w^2(x,p)\frac{P}{1+P}dp= 0.
\end{equation}
Using the decomposition $\tilde{f}= (a|p|^2+\tilde{b}p_x)(1+P)+w$ of $\tilde{f}$ into its hydrodynamic and non hydrodynamic components, and setting
\begin{eqnarray*}
a_\infty = \lim _{x\rightarrow +\infty }a(x),\quad \tilde{b}_\infty = \lim _{x\rightarrow +\infty }\tilde{b}(x),
\end{eqnarray*}
it follows from (\ref{asympt-uniq1})-(\ref{asympt-uniq2}) that
\begin{eqnarray*}
\lim _{x\rightarrow +\infty }\int p_x\Big( (a(x)|p|^2+\tilde{b}(x)p_x)(1+P)\Big) ^2\frac{P}{1+P}dp= 0,\quad \text{i.e.}\quad a_\infty \tilde{b}_\infty = 0.
\end{eqnarray*}
\hspace*{0.2in}\\
Below this will be improved to (1.10) $\tilde{b}_{\infty}=0$.\\
\hspace{1cm}\\
But first we prove the existence of a solution $\tilde{f}\in D$ to the Milne problem (\ref{milne2})-(\ref{milne-bc2})-(\ref{energy-flow2}). It will be obtained  as the limit when $l\rightarrow +\infty $ of the sequence $(\tilde{f}_l)_{l\in \N ^*}$ of solutions to the stationary linearized equation on the slab $ [ 0,l ] $ with specular reflection at $x=l$, i.e.
\begin{equation}\label{milne-specular1}
p_x\partial _x\tilde{f}_l= L\tilde{f}_l, \quad x\in [ 0,l] ,\hspace*{0.05in}p_x\in \R  ,\hspace*{0.05in}p_r\in \R ^+,
\end{equation}
\begin{equation}\label{milne-specular2}
\tilde{f}_l(0,p)= \tilde{f}_0(p),\quad p_x>0,
\end{equation}
\begin{equation}\label{milne-specular3}
\tilde{f}_l(l,p_x,p_r)= \tilde{f}_l(l,-p_x,p_r),\quad p_x<0.
\end{equation}
Switch from given in-data and no inhomogeneous term, to zero indata and an inhomogeneous term.
Let $\epsilon >0$ be given. Let the subspace $D(A)$ of $L^2_{p_r(1+\lvert p\rvert )^3\frac{P}{1+P}}((0,l)\times \R\times \R^+)$ be defined by
\[ \begin{aligned}
D(A)= \{ &g\in L^2_{p_r(1+\lvert p\rvert )^3\frac{P}{1+P}}((0,l)\times \R\times \R^+); \hspace*{0.03in}p_x\partial _xg\in L^2_{p_r(1+\lvert p\rvert )^{3}\frac{P}{1+P}}((0,l)\times \R\times \R^+),\\
&\hspace*{0.03in}g(0,p)= 0, p_x>0,\quad g(l,p_x,p_r)= g(l,-p_x,p_r), p_x<0\} .
\end{aligned}\]
The operator $A$ defined on $D(A)$ by
\begin{eqnarray*}
(Ag)(x,p)= \epsilon g(x,p)+p_x\partial _xg(x,p)
\end{eqnarray*}
is $m$-accretive since $I-\frac{1}{2\epsilon }A$ is bijective. Indeed, for any $f\in L^2_{p_r(1+\lvert p\rvert )^3\frac{P}{1+P}}((0,l)\times \R\times \R^+)$ , there is a unique $g\in D(A)$ such that
\begin{equation}\label{accret1}
(I-\frac{1}{2\epsilon }A)g= f\quad \text{i.e.}\quad \frac{1}{2}g-\frac{1}{2\epsilon }p_x\partial _xg= f.
\end{equation}
Here $g$ is explicitly given by
\[ \begin{aligned}
g(x,p)&= -\frac{2\epsilon }{p_x}\int _0^xf(y,p)e^{\epsilon \frac{x-y}{p_x}}dy,\quad p_x>0,\\
g(x,p)&= \frac{2\epsilon }{p_x}\Big( \int _0^lf(y,p)e^{\epsilon \frac{x+y-2l}{p_x}}dy+\int _x^lf(y,p)e^{\epsilon \frac{x-y}{p_x}}dy\Big) ,\quad p_x<0.
\end{aligned}\]
It belongs to $L^2_{p_r(1+\lvert p\rvert )^3\frac{P}{1+P}}((0,l)\times \R\times \R^+)$ since multiplying (\ref{accret1}) by $2g(1+\lvert p\rvert )^3\frac{P}{1+P}$, then integrating on $[ 0,l] \times \R ^3$ implies that
\[ \begin{aligned}
\int g^2(x,p)(1+\lvert p\rvert )^3\frac{P}{1+P}dxdp&+\frac{1}{2\epsilon }\int _{p_x<0}\lvert p_x\rvert (1+\lvert p\rvert )^3g^2(0,p)\frac{P}{1+P}dp\\
&= 2\int f(x,p)g(x,p)(1+\lvert p\rvert )^3\frac{P}{1+P}dp\\
&\leq \int f^2(x,p)(1+\lvert p\rvert )^3\frac{P}{1+P}dxdp+\int g^2(x,p)(1+\lvert p\rvert )^3\frac{P}{1+P}dxdp.
\end{aligned}\]
It then follows from (\ref{accret1}) that $p_x\partial _xg\in L^2_{p_r(1+\lvert p\rvert )^3\frac{P}{1+P}}((0,l)\times \R\times \R^+)$.\\
Since $-L$ is an accretive operator, from here by an $m$-accretive study of $A-L$, there exists a solution \\
$\tilde{f}_\epsilon \in L^2_{p_r(1+\lvert p\rvert )^3\frac{P}{1+P}}((0,l)\times \R\times \R^+)$ to
\begin{equation}\label{approx-milne}
\epsilon \tilde{f}_\epsilon +p_x\partial _x\tilde{f}_\epsilon = L\tilde{f}_\epsilon , \quad x>0,\hspace*{0.05in}p_x\in \R,\hspace*{0.05in}p_r\in \R ^+,
\end{equation}
\[ \begin{aligned}
&\tilde{f}_\epsilon (0,p)= \tilde{f}_0(p),\quad p_x>0,\\
&\tilde{f}_\epsilon (l,p_x,p_r)= \tilde{f}_\epsilon (l,-p_x,p_r),\quad p_x<0.
\end{aligned}\]
In order to prove that there is a converging subsequence of $(\tilde{f}_\epsilon )$ when $\epsilon $ tends to zero, split $\tilde{f}_\epsilon $ into its hydrodynamic and non-hydrodynamic parts as
\begin{eqnarray*}
\tilde{f}_\epsilon (x,p)= (a_\epsilon (x)|p|^2+b_\epsilon (x)p_x)(1+P)+w_\epsilon (x,p),
\end{eqnarray*}
with
\begin{equation}\label{orthog-conditions}
\int p_xw_\epsilon Pdp= \int |p|^2w_\epsilon Pdp= 0.
\end{equation}
Multiply (\ref{approx-milne}) by $\tilde{f}_\epsilon \frac{P}{1+P}$, integrate w.r.t. $(x,p)\in [ 0,l] \times \R \times \R ^+$ and use the spectral inequality (\ref{spectral-in}), so that $(w_\epsilon )$ is uniformly bounded in $L^2_{p_r(1+\lvert p\rvert )^3\frac{P}{1+P}}([ 0,l] \times \R \times \R ^+)$. Notice that the boundary term at $l$ vanishes. And so, up to a subsequence, $(w_{\epsilon })$ weakly converges in $L^2_{p_r(1+\lvert p\rvert )^3\frac{P}{1+P}}([ 0,l] \times \R \times \R ^+))$ to some function $w$. Moreover, the same argument as for getting (\ref{pointwise-exp-w-decay}) can be used here, so that
\begin{equation}\label{pointwise-exp-w-eps-decay}
e^{2\eta x}\int  (1+|p|)^3w_{\epsilon }^2(x,p)\frac{P}{1+P}dp\leq c,\quad x\in [ 0,l] .
\end{equation}
Expressing $\int _{p_x>0}p_x\tilde{f}_\epsilon (0,p)\frac{P}{1+P}dp$ (resp. $\int _{p_x>0}p_x|p|^2\tilde{f}_\epsilon (0,p)\frac{P}{1+P}dp$) in terms of $a_\epsilon (0)$, $b_\epsilon (0)$ and $w_\epsilon (0,\cdot )$ leads to
\[ \begin{aligned}
&a_{\epsilon }(0)\int _{p_x>0}p_x|p|^2Pdp+b_{\epsilon }(0)\int _{p_x>0}p_x^2Pdp\\
&= \int _{p_x>0}p_x\tilde{f}_0(p)\frac{P}{1+P}dp-\int _{p_x>0}p_xw_\epsilon (0,p)\frac{P}{1+P}dp,
\end{aligned}\]
and
\[ \begin{aligned}
&a_{\epsilon }(0)\int _{p_x>0}p_x|p|^4Pdp+b_{\epsilon }(0)\int _{p_x>0}p_x^2|p|^2Pdp\\
&= \int _{p_x>0}p_x|p|^2\tilde{f}_0(p)\frac{P}{1+P}dp-\int _{p_x>0}p_x|p|^2w_\epsilon (0,p)\frac{P}{1+P}dp.
\end{aligned}\]
By the Cauchy-Schwartz inequality and (\ref{pointwise-exp-w-eps-decay}) taken at $x= 0$, it follows that
\begin{eqnarray*}
\int _{p_x>0}p_xw_\epsilon (0,p)\frac{P}{1+P}dp\quad \text{and}\quad \int _{p_x>0}p_x|p|^2w_\epsilon (0,p)\frac{P}{1+P}dp
\end{eqnarray*}
are bounded. Consequently, $(a_\epsilon (0) ,b_\epsilon (0))$ is uniformly bounded with respect to $\epsilon $.
Moreover, $f_\epsilon $ solves (\ref{approx-milne}) if and only if
\begin{eqnarray}
\epsilon \Big( (a_\epsilon |p|^2+b_\epsilon p_x)(1+P)+w_\epsilon \Big) +(a'_\epsilon p_x|p|^2+b'_\epsilon p_x^2)(1+P)+p_x\partial _xw_\epsilon = Lw_\epsilon .
\end{eqnarray}
Multiplying the previous equation by $p_xP$ (resp. $(p^2+n)P$) and integrating w.r.t. $p$, implies that
\[ \begin{aligned}
&\epsilon b_\epsilon \int p_x^2P(1+P)dp+\gamma a'_\epsilon +\Big( \int p_x^2w_\epsilon Pdp\Big) '= 0,\\
&\epsilon a_\epsilon \int |p|^4P(1+P)dp+\gamma b'_\epsilon +\Big( \int p_x|p|^2w_\epsilon Pdp\Big) '= 0.
\end{aligned}\]
Consequently, denoting by
\begin{eqnarray*}
\alpha =\sqrt{\int p_x^2P(1+P)dp}\quad \text{and}\quad \beta =\sqrt{\int |p|^4P(1+P)dp},
\end{eqnarray*}
it holds that
\[ \begin{aligned}
\gamma a_\epsilon (x)&= -\int p_x^2w_\epsilon (x,p)Pdp+(\gamma a_\epsilon (0)+\int p_x^2w_\epsilon (0,p)Pdp)e^{\frac{\alpha \beta \epsilon }{\gamma }x}\\
&+\epsilon \int _0^x\Big( \int p_x\frac{\alpha ^2}{\gamma }(p^2+n)w_\epsilon (y,p)Pdp\Big) e^{\frac{\alpha \beta \epsilon }{\gamma }(x-y)}dy,\quad x\in [ 0,l] ,\\
\gamma b_\epsilon (x)&= -\int p_x|p|^2w_\epsilon (x,p)Pdp+(\gamma b_\epsilon (0)+\int p_x|p|^2w_\epsilon (0,p)Pdp)e^{-\frac{\alpha \beta \epsilon }{\gamma }x}\\
&+\epsilon \int _0^x\Big( \int \frac{\beta ^2}{\gamma }p_x^2w_\epsilon (y,p)Pdp\Big) e^{-\frac{\alpha \beta \epsilon }{\gamma }(x-y)}dy,\quad x\in [ 0,l] .
\end{aligned}\]
Together with the bounds of $(a_{\epsilon }(0),b_{\epsilon }(0))$ and (\ref{pointwise-exp-w-eps-decay}), this implies that $(a_\epsilon )$ (resp. $(b_\epsilon )$) is bounded in $L^2$. And so, up to a subsequence, $\tilde{f}_\epsilon $ weakly converges in $L^2_{p_r(1+\lvert p\rvert )^3\frac{P}{1+P}}((0,l)\times \R\times \R^+)$ to a solution $\tilde{f}_l$ of (\ref{milne-specular1})-(\ref{milne-specular2})-(\ref{milne-specular3}).\\
Similar arguments can be used in order to prove that up to a subsequence, $(\tilde{f}_l)$ converges  to a solution $\tilde{f}$ of the Milne problem (\ref{milne2})-(\ref{milne-bc2})-(\ref{energy-flow2}) when $l$ tends to $+\infty $. Indeed, if $\tilde{f}_l$ admits the decomposition
\begin{eqnarray*}
\tilde{f}_l(x,p)= (a_l(x)|p|^2+b_l(x)p_x)(1+P)+w_l(x,p),
\end{eqnarray*}
with
\begin{eqnarray*}
\int p_xw_lPdp= \int |p|^2w_lPdp= 0,
\end{eqnarray*}
then the sequence $(w_l)$ is bounded in $L^2_{p_r(1+\lvert p\rvert )^3\frac{P}{1+P}}(\R^+\times \R \times \R^+)$ and pointwise in $x$ as in (\ref{pointwise-exp-w-eps-decay}). And so, up to a subsequence, $(w_l)$ converges weakly in $L^2_{p_r(1+\lvert p\rvert )^3\frac{P}{1+P}}(\R^+\times \R\times \R^+)$ and also weak star in $x$, weak in $p$ in $L^\infty (\R^+;L^2_{p_r(1+\lvert p\rvert )^3}(\R\times \R^+))$. The sequences $(a_l)$ and $(b_l)$ satisfy
\[ \begin{aligned}
&\Big( \gamma a_l +\int p_x^2w_l Pdp\Big) ^\prime = 0,\quad \Big( \gamma b_l +\int p_x|p|^2w_l Pdp\Big) ^\prime = 0,
\end{aligned}\]
so that
\[ \begin{aligned}
&\gamma a_l(x)= -\int p_x^2w_l(x,p) Pdp+\gamma a_l(0)+\int p_x^2w_l(0,p) Pdp,\\
&\gamma b_l(x)= -\int p_x|p|^2w_l(x,p) Pdp+\gamma b_l(0)+\int p_x|p|^2w_l(0,p) Pdp.
\end{aligned}\]
It follows that the sequences $(a_l)$ and $(b_l)$ are uniformly bounded on $\R^+$, and so, up to a subsequence, converge weak star in $x$. The limit of $(\tilde{f}_l)$ is a weak solution to the problem. This weak solution belongs to $D$.\\
\hspace*{0.2in}\\
We can now prove that $\tilde{b}_{\infty}=0$. For this we notice that the discussion of this section up to (3.11) included, also holds for $\tilde{f}_l$, $W$ being nonnegative on $[ 0,l] $ because it is non increasing and vanishes at $l$. The discussion from (3.12) leading up to (3.15) is valid as well.
But for $f_l$ it holds that $\tilde{b}_l(l)=0$, and so (\ref{asympt-b}) taken at $x= l$ leads to $|\tilde{b}_{l\infty}|\leq ce^{-\eta l}$.\\
Take $\beta \geq\alpha\gg0$. Using (\ref{asympt-b}) again implies that for all $l>\beta$,
\begin{eqnarray*}
|\tilde{b}_l(x)|\leq|\tilde{b}_l(x)-\tilde{b}_{l\infty}|+ce^{-\eta\alpha}\leq2ce^{-\eta\alpha},\quad x\geq \alpha .
\end{eqnarray*}
It follows that
\begin{eqnarray*}
\lvert \tilde{b}(x)\rvert \leq2ce^{-\eta\alpha},\quad x\geq\alpha.
\end{eqnarray*}
Hence
\begin{eqnarray*}
\lim_{x\rightarrow\infty}\tilde{b}(x)=0=\tilde{b}_{\infty}.
\end{eqnarray*}
%
%
\hspace{1cm}\\
The uniqueness of the solution of the Milne problem (\ref{milne2})-(\ref{milne-bc2})-(\ref{energy-flow2})  can be proven as follows. Let $\tilde{f}\in D$ be solution to the Milne problem (\ref{milne2})-(\ref{milne-bc2})-(\ref{energy-flow2}) with zero indatum at $x= 0$ and zero energy flow. Let
\begin{eqnarray*}
\tilde{f}(x,p)= a(x)|p|^2(1+P)+b(x)p_x(1+P)+w(x,p)
\end{eqnarray*}
be its orthogonal decomposition. By (\ref{exp-decrease-W})
\begin{equation}\label{asympt-pf-uniq1}
\lim _{x\rightarrow +\infty }\int p_x\tilde{f}^2(x,p)\frac{P}{1+P}dp= 0.
\end{equation}
Multiply the equation
\begin{equation}\label{milne-2}
p_x\partial _x\tilde{f}= L\tilde{f},
\end{equation}
by $\tilde{f}\frac{P}{1+P}$, integrate over $ ] 0,+\infty [\times \R ^3 $ and use the spectral inequality. Then,
\[ \begin{aligned}
\frac{1}{2}\int _{p_x<0}\lvert p_x\rvert \tilde{f}^2(0,p)\frac{P}{1+P}dp+\nu _0\int _0^{+\infty }\int w^2(x,p)\frac{P}{1+P}dpdx&\leq -\frac{1}{2}\lim _{x\rightarrow +\infty }\int  p_x\tilde{f}^2(x,p)\frac{P}{1+P}dp\\
&= 0.
\end{aligned}\]
And so,
\begin{eqnarray*}
\tilde{f}(0,\cdot )= 0,\quad w(\cdot ,\cdot )= 0.
\end{eqnarray*}
Equation (\ref{milne-2}) reduces to
\begin{eqnarray*}
\partial _x\tilde{f}= 0,
\end{eqnarray*}
so that together with $\tilde{f}(0,\cdot )= 0$, it holds that $a(\cdot )= b(\cdot )= 0$. Hence $\tilde{f}$ is identically zero.\\
\hspace*{0.2in}\\
\hspace*{0.2in}\\
{\bf Acknowledgement.} Useful referee comments on the mathematical physics aspects, have helped
to improve the paper.
\hspace*{0.2in}\\
\hspace{1cm}\\
%
%
%
%
%
%
%


\begin{thebibliography}{99}
%
\bibitem
{AN1} L. Arkeryd, A. Nouri, {\it Bose condensates in interaction with excitations - a kinetic model}, Commun. Math. Phys. 310 Issue 3 765-788 (2012).


\bibitem
{AN3} L. Arkeryd, A. Nouri, {\it Bose condensates in interaction with excitations - a two-component, space dependent model close to equilibrium},  in preparation.


\bibitem
{AN2} L. Arkeryd, A. Nouri, {\it On the Milne problem and the hydrodynamic limit for a steady Boltzmann equation model}, J. Stat. Phys. 99  993-1019 (2000).


\bibitem
{BB} N. Bernhoff, A. V. Bobylev, {\it Discrete velocity models and dynamic systems}, Lecture Notes on the discretization of the Boltzmann equation, World Sci. Pub. I 203-222 (2003).

\bibitem
{BCN} C. Bardos, R.E. Caflish, B. Nicolaenko, {\it The Milne and Kramers problems for the Boltzmann equation of a hard sphere gas}, Commun. Pure Appl. Math. Vol XXXIX 323-352 (1986).

\bibitem
{BGS} C. Bardos, F. Golse, Y. Sone, {\it Half-space problems for the Boltzmann equation: a survey}, J. Stat. Phys. 124 Issue 2-4 275-300 (2006).

\bibitem
{BT} A. V. Bobylev, G. Toscani, {\it Two-dimensional half space problems for the Broadwell discrete velocity model}, Contin. Mech. Thermodyn. 8 257-274 (1996).

\bibitem
{C} C. Cercignani, {\it Half-space problems in the kinetic theory of gases}, Lect. Notes Phys.
249 35-51, Springer-Verlag New-York (1987).

\bibitem
{CGS} F. Coron, F. Golse, C. Sulem, {\it A classification of well-posed kinetic layer problems}, Commun. Pure Appl. Math. 41 Issue 4 409-435 (1988).

\bibitem
{CME} C. Cercignani, R. Marra, R. Esposito, {\it The Milne problem with a force term}, Transport Theory and Statistical Physics 27 1-33 (1998).

\bibitem
{GP} F. Golse, F. Poupaud, {\it Stationary solutions of the linearized Boltzmann equation in a half-space}, Math. Methods Appl. Sci. 11 Issue 4 483-502 (1989).

\bibitem
{M1} N. Maslova, {\it Kramers problems in the kinetic theory of gases}, USSR Comput. Math. Phys. 22 208-219 (1982).

\bibitem
{M2} N. Maslova, Nonlinear Evolution Equations, World Scientific, Singapore 1993.

\bibitem
{P} F. Poupaud, {\it Diffusion approximation of the linear semiconductor equation: analysis of boundary layers}, Asymptotic Analysis 4 293-317 (1991).

\bibitem
{So1} Y. Sone, Kinetic Theory and Fluid Dynamics, Birkhauser, Boston 2002.


\bibitem
{So2} Y. Sone, Molecular Gas Dynamics, Birkhauser, Boston 2006.


\bibitem
{UYY1} S. Ukai, T. Yang, S.-H. Yu, {\it Nonlinear boundary layers of the Boltzmann equation: I. Existence}, Commun. Math. Phys. 236 373-393 (2003).

\bibitem
{UYY2} S. Ukai, T. Yang, S.-H. Yu, {\it Nonlinear stability of boundary layers of the Boltzmann equation; I. The case $\mathcal{M}_\infty <-1$}, Commun. Math. Phys. 244 99-109 (2003).

\end{thebibliography}
\end{document}